%% file: main.tex
\documentclass[sigconf]{acmart}
\setcopyright{none} 
\renewcommand\footnotetextcopyrightpermission[1]{} 
\AtBeginDocument{%
  }

\usepackage{graphicx}
\usepackage{amsmath}
\usepackage{csquotes}
\usepackage[draft,inline,nomargin,index]{fixme}
\usepackage{multirow}
\usepackage{nicefrac}
\usepackage{siunitx}
\usepackage{array,framed}
\usepackage{subcaption}
\usepackage{tabularx}
\usepackage[flushleft]{threeparttable}

\usepackage{
  color,
  float,
  epsfig,
  wrapfig,
  graphics,
  graphicx,
  caption}
\graphicspath{{./images/}}
\DeclareGraphicsExtensions{.pdf,.jpeg,.png}

\usepackage{textcomp}
\usepackage{setspace}
\usepackage{latexsym,fancyhdr}

\usepackage{enumerate}
\usepackage{paralist}

\usepackage[inline]{enumitem}
\usepackage[linesnumbered,ruled,vlined]{algorithm2e}

\usepackage{algpseudocode}
\usepackage{graphics}
\usepackage{csvsimple}
\usepackage{balance}
\usepackage{mathtools}

\usepackage{amsmath}

\usepackage{balance}  
\usepackage{listings}
\usepackage{upquote}
\usepackage{xcolor}
\usepackage{url}  
\usepackage{url}

\usepackage{paralist}
\usepackage{ragged2e}
\usepackage{fixltx2e}
\usepackage{tikz}
\usepackage{mdframed}
\usepackage{makecell}

\mdfsetup{skipabove=\topskip,skipbelow=\topskip}
\newcounter{prompt}[section]

\input{sources/defs}

 \fxsetup{
 status=draft,
 theme=color,
 mode=multiuser
 }

\FXRegisterAuthor{j}{ejulian}{\color{blue}JS}
\FXRegisterAuthor{s}{Shriti}{\color{purple}SP}

\definecolor{mygreen}{rgb}{0,0.4,0}
\definecolor{myblue}{rgb}{0.0,0.1,0.3}
\definecolor{mycodeblue}{rgb}{0.0,0.28,0.67}
\definecolor{myred}{rgb}{0.7,0.0,0.2}
 
\lstdefinelanguage{REST}
{ 
morekeywords={GET, POST, PUT, DELETE},
sensitive=false,
morecomment=[l]{\/\*\*},
commentstyle=\color{mygreen},  	
keepspaces=true,                 	
keywordstyle=\color{myblue}\bfseries,       	
morestring=[b]'
}

\lstdefinelanguage{mySQL}
{
language=SQL,
commentstyle=\color{mygreen},  	
keywordstyle=\color{myblue}\bfseries,       	
}

\lstdefinelanguage{go2}{
  sensitive = true,
  language=java,
  keywords = [2]{interface, type, bool, error, if, else},
  commentstyle=\color{mygreen},  	
  keywordstyle=\color{mycodeblue}\bfseries,       	
  keywordstyle=[2]\color{mycodeblue}\bfseries       	
}

\lstdefinelanguage{codeblue}{
  basicstyle=\color{blue}\ttfamily\small  
  breaklines=true, 
  breakatwhitespace=true, 
  keywordstyle=\color{mycodeblue}, 
}

\lstdefinelanguage{prompt}
{ 
sensitive=false,
morecomment=[l]{\/\*\*},
commentstyle=\color{mygreen},  	
keepspaces=true,                 	
keywordstyle=\color{myblue}\bfseries,       	
morestring=[b]'
}

\begin{document}

\title{\Large \bf \sysname{:} A Policy Framework for Securing Cloud APIs by Combining Application Context with Generative AI}

\author{Shriti Priya}
\authornote{Equal contribution from both authors}
\affiliation{%
  \institution{IBM Research}
  \city{Yorktown Heights}
  \state{NY}
  \country{USA}
}
\author{Julian James Stephen}
\affiliation{%
  \institution{IBM Research}
  \city{Yorktown Heights}
  \state{NY}
  \country{USA}
}
\authornotemark[1]
\author{Arjun Natarajan}
\affiliation{%
  \institution{IBM Research}
  \city{Yorktown Heights}
  \state{NY}
  \country{USA}
}



\input{sources/abstract}
\maketitle
\pagestyle{empty}
\input{sources/introduction}

\input{sources/background}

\input{sources/design}
\input{sources/detail}

\input{sources/prompt_tuning}

\input{sources/implementation}

\input{sources/evaluation}
\input{sources/related}
\input{sources/conclusion}



\bibliographystyle{ACM-Reference-Format}
\bibliography{bibtex/cloud.bib, bibtex/policy.bib, bibtex/AI.bib, bibtex/threats.bib}


\end{document}

%% file: sources/defs.tex
\usepackage{xparse}
\newcommand{\secref}[1]{\mbox{Section~\ref{#1}}}

\newcommand{\figref}[1]{\mbox{Figure~\ref{#1}}}
\newcommand{\lstref}[1]{\mbox{Listing~\ref{#1}}}

\renewcommand{\eqref}[1]{\mbox{Equation~(\ref{#1})}}

\newcommand{\tabref}[1]{\mbox{Table~\ref{#1}}}

\newcommand{\Modeone}{Single mode}
\newcommand{\modetwo}{parallel mode}
\newcommand{\Modetwo}{Parallel mode}

\newcommand{\dsone}{top 25 APIs }

\newcommand{\Dsone}{Top 25 APIs }
\newcommand{\Dstwo}{Finance APIs}
\newcommand{\Dsthree}{E-commerce APIs}

\newcommand{\Login}{Login}
\newcommand{\Logout}{Logout}
\newcommand{\Fileupload}{File upload}
\newcommand{\Userregistration}{User registration}
\newcommand{\Commenting}{Commenting}
\newcommand{\Purchaseproduct}{Purchase product}
\newcommand{\Addtocart}{Add to cart}
\newcommand{\Responsedatalimit}{Response data limit}
\newcommand{\responsedatalimit}{response data limit}

\newcommand{\Containsauthorizationtokens}{Contains authorization tokens}
\newcommand{\containsauthorizationtokens}{contains authorization tokens}

\newcommand{\sysname}{Paladin} 

\SetCommentSty{mycommfont}

\newcommand{\nip}[1]{\vspace{1ex}\noindent\textbf{#1}}

\newcommand{\mathnip}[1]{\vspace{1ex}\noindent{#1}}

\newcolumntype{L}{>{\centering\arraybackslash}m{3cm}}

\makeatletter
\newcommand\newtag[2]{#1\def\@currentlabel{#1}\label{#2}}
\makeatother

%% file: sources/abstract.tex
\begin{abstract}
    Enterprises and organizations today increasingly deploy in-house, cloud
    based applications and APIs for internal operations or external customers.
    These deployments deal with increasing number of threats, despite security
    features offered by cloud service providers. This work focus on threats that
    exploit application layer vulnerabilities of cloud workloads. Prevention and
    mitigation measures against such threats need to be cognizant of application
    semantics, posing a hurdle to existing solutions. 

In this work, we design and implement a security framework that allow cloud
workload administrators to easily define and enforce policies capable of
preventing  \begin{enumerate*}[label=(\roman*)]
\item unrestricted resource consumption, 
\item unrestricted access to sensitive business flows, and
\item broken authentication.
\end{enumerate*}
 Our framework, \sysname{,}  leverages large language models to extract
 sufficient semantic meaning from API requests to provide cloud administrators
 with an application agnostic policy definition interface. Once defined,
 requests are automatically matched with relevant policies and enforced by high
 performance proxies. 
 Evaluations with our prototype show that such a framework has broad
 applicability across applications, good policy identification accuracy
 (81\%), and reasonable overheads (14\%), making it   
  substantially easier to define and enforce cross
 application policies.

\end{abstract}

%% file: sources/introduction.tex
\section{Introduction\label{sec:introduction}} 
Institutions, industries and governments utilize cloud based compute resources
in one form or another. Today, cloud applications encompass banking
systems~\cite{DBLP:journals/clsr/HonM18}, space
research~\cite{DBLP:conf/iccnc/ShackelfordW12}, health
care~\cite{DBLP:journals/ijinfoman/Sultan14}, and other key services in society.
Within web based development methodologies, creating application programming
interfaces (APIs), especially RESTful~\cite{masse2011rest} APIs have become
extremely popular. Ensuring these applications or APIs are secure from attacks,
data leaks and other misuse, while remaining available for its intended use is
of primary concern for these organizations.

Layer 7 attacks target application layer (OSI~\cite{OSI}) resources like web
APIs and data servers. Researchers have extensively studied how to prevent
various layer 7 attacks like denial of service
(DoS)~\cite{dosSensor,dosNeedham,ddos1}, injection
attacks~\cite{injection1,codeinjection,sqlinjection}, cross-site request forgery
(CSRF)~\cite{csrf1,zeller2008cross} and more.  Still, applications and APIs are
getting deployed with known and preventable vulnerabilities. This leads to an
orthogonal need for organizations to protect their many and varied cloud based
APIs with a strong set of policies. The primary goal of these set of policies
would be to ensure an organizational level of protection, irrespective of the
quality or safety of individual applications.

\nip{Challenges.}  Towards defining and enforcing such policies, organizations
    face a wide array of  challenges. Consider a scenario where applications in
    the cluster are exposing endpoints which are not secured against
    unrestricted resource consumption. E.g., an API endpoint that returns a
    certain number of records in its response, based on an input HTTP parameter,
    like  {\RaggedRight \lstinline[language=REST, columns=fullflexible,
    breaklines=true]!GET /query?numResults=10!. } In this example, the web
    server is supposed to return  \lstinline{10} records in its response. But,
    if the application does not impose a maximum limit for the
    \lstinline{numResults} parameter, and the query result size is high, an
    attacker can make the same request with  \lstinline{numResults=100000},
    which can paralyze the web server. A second application may have a similar
    endpoint but with a different parameter name. E.g.,
    \lstinline[language=REST,breaklines=true]{GET /search?count=10}, with the
    parameter \lstinline{count} now controlling the number of records in the
    response. Across multiple applications, such differences can be over the
    name of the parameter, semantics of the end point (an array of objects
    instead of search results), HTTP request type (e.g., POST vs GET) etc. To
    define and enforce a cluster wide policy across such differences, security
    administrators or policy writers need to rely on API developers to provide
    prior information about their APIs. The administrators then need to
    understand the application semantics and nuances of different API groups to
    secure their clusters effectively. Further, as API endpoints and API
    parameters change, these policies need to be updated. As the number of
    applications and APIs in the cloud increase, the burden on
    cloud security administrators becomes untenable. 

\nip{Solution outline.} In this work we design a security framework (\sysname{)}
that allow cloud security administrators to easily define and enforce cross
application policies. We show that such policies are effective for  preventing a
wide array of security risks including:
\begin{enumerate}[label=\textbf{T.\arabic*},ref=T.\arabic*,align=left]
    \item \label{rc} Unrestricted resource consumption
    \item \label{sbf} Unrestricted access to sensitive business flows, and
    \item \label{ba} Broken authentication
\end{enumerate} 
 These categories are more formally defined in~\secref{sec:tm}. In general,
these threats originate from  applications not performing proper boundary
checks, authorizations, and validations on each request. Our framework,
\sysname{,}  provides an application and API independent abstraction to define
policies that perform such checks. E.g., consider the HTTP request {\RaggedRight
\lstinline[language=REST, breaklines=true]!GET v1/query?numResults=10!} from
before. A policy author can specify that all requests with HTTP parameters
having a \emph{similar} meaning to \lstinline{numResults} should have 
parameter values bound within
a predefined variance. \sysname{} then utilizes large language
models~\cite{Bommasani2021FoundationModels} to identify HTTP parameters that
have the same semantics as \lstinline{numResults}. These may include
\lstinline{size}, \lstinline{num_results}, \lstinline{count}, etc. More complex
policies rely on identifying semantics of the complete request, not just the
request parameters. E.g., both \lstinline[language=REST]{GET /query} and
\lstinline[language=REST]!GET /search! are requests resulting in the server
returning a list of records in its response. 
\sysname{} automates the 
job of understanding and grouping similar APIs from  different applications, 
allowing cloud administrators to define policies much faster and easier.
\sysname{} also curates and utilizes past
context of requests to enable richer policies.

\nip{Key contributions.} Many common applications expose endpoints
potentially susceptible to the types of attacks listed above 
(\ref{rc}, \ref{sbf} and \ref{ba})~\cite{owaspapitop10}. We observe that
appropriately tuned large language models can be leveraged to successfully
extract sufficient semantic meanings from HTTP API requests to create a library
of effective policies. Our work is not a replacement for any existing security
mechanisms (firewalls, gateways, intrusion detection systems, etc.), but rather 
a new control instrument for cloud administrators.
Towards this, our key contributions are:
\begin{enumerate}
    \item A cloud platform level framework that allow administrators to define
    cross application API policies to protect against specific
    layer-7  threats
    \item A novel design that combines request and environment context with 
    the intent of the request,  making policies effective
    \item Leverage large language models to group and extract relevant information 
    from API requests to enforce relevant policies
    \item Evaluations  showing broad effectiveness  of the framework as a practical solution

\end{enumerate}

 The remainder of this paper is organized as follows. We present background
 information on relevant systems and threat model 
 in \secref{sec:background}. We provide details of our framework in
 \secref{sec:design}. In \secref{sec:policydef},  we discuss how to define policies to prevent the identified 
 risks.  \secref{sec:prompts} describe how we interface with the 
 large language model and \secref{sec:impl} describe our implementation.   
 We present our evaluation results in \secref{sec:eval}.
 \secref{sec:related} contrasts with related work, and
 \secref{sec:conclusion} concludes with final remarks.

%% file: sources/background.tex
\section{Background\label{sec:background} and Threat Model} 
This section describe background information and 
 threat model  of  our framework.

\subsection{Cloud architecture}


\nip{Containers and orchestration.}  Containers~\cite{containersSolteszPFBP07}
are operating system constructs that provide lightweight virtualization.
Containers allow multiple user space instances to share the same host OS while
allowing multiple applications to run in isolation and in parallel. Container
orchestration defines how to select, deploy, monitor, scale, and control the
interactions of multi-container applications in the cloud. Open-source container
orchestration platforms like Kubernetes~\cite{k8} make up the control plane of
typical cloud environments. Using an orchestration platform greatly improves
ease of use and provide improved security benefits~\cite{k8security}. 

\nip{Service and data proxies.} In a container orchestration platform like
Kubernetes, deployed application micro services (e.g., Pods in
Kubernetes~\cite{k8pod}) are meant to be ephemeral. Applications are usually
assigned a  more permanent \emph{service} address which automatically reroutes
traffic to the changing application IP (pod IP)  address. In other words, 
service addresses are long lived and used by micro services to talk to one
another. A proxy acts as an intermediary between these service to
service calls or client to service calls. Similar to the service proxies, a data proxy sits in between an
application and a data store. Proxies are useful for providing load balancing,
monitoring and security for the service they are proxying.

\nip{REST APIs, HTTP and HTTPS.}
Application programming interfaces (APIs) are mechanisms that allow different
cloud (or any software) components to communicate with each other. APIs may use
a predefined set of protocol for this communication, HTTP~\cite{rfc2616} and
HTTPS~\cite{rfc2818} being the most popular for cloud based API ecosystems. 
REST~\cite{fielding2000architectural} is a protocol-agnostic architecture 
for APIs, with constraints like communication being client-server, stateless, etc. 
In this paper, we refer to a group of APIs (irrespective of the architecture
or underlying protocol),  providing functionality for one
application simply as an application. 

\nip{Layer-7 policies.} The service plane of a container orchestration platform
provide the ability to define policies that can be plugged in various stages of
an application's runtime. Service layer policies enable policy enforcement
points in front of services. These policies allow security administrators to
ensure containers and container traffic meets their prerequisite conditions.

\nip{Large Language Models.}
Large Language Models (LLMs) are AI models  trained on very broad data and at
scale, making it feasible to adapt them to a wide range of  tasks. They are
based on standard deep learning~\cite{Brown,Bommasani2021FoundationModels} and
transfer learning~\cite{Thrun1998} but they have been shown to be effective
across many tasks. 

\subsection{Threat Model \label{sec:tm}}
For our work, we consider multiple application workloads deployed in a
cloud cluster, each exposing a set of  APIs.
Applications in this environment may not be following security best
practices and may have known vulnerabilities. 
Attackers may invoke application APIs as many times as needed
with the following goals and capabilities. 

\nip{Unrestricted resource consumption (\ref{rc}).} In this case,
the attacker sends multiple requests to the API server with the intention to
 starve server resources. APIs can be vulnerable if they do not limit 
 file upload sizes, number of records per page to return, maximum number of 
 processor or file descriptors, etc. 

\nip{Unrestricted access to sensitive business flows (\ref{sbf}).} Attacker will try to find
sensitive business flows and manually or by automating access to these flows,
try to cause harm to the business. This can include denial of
inventory~\cite{OWASPDenialOfInventory}, scalping, spamming and more. 

\nip{Broken authentication.  (\ref{ba})} Attackers may try to gain control over
application user's accounts. Attackers will then utilize this control to view
the user's personal data and perform actions on the user's behalf. Attackers may
try to brute force a user account by trying different authorization mechanisms.
Further examples of an attacker's intention and the corresponding threat
category is given in  \tabref{tab:threats}.

\begin{table*}[ht]
    \caption{\label{tab:threats} Attacker intentions}
    \begin{tabularx}{\linewidth}{lX}
    \hline
    Intention &  Definition \\ \hline
    \makecell{Server resource limit \\(\ref{rc})} & Make the server retrieve maximum
    possible records in it's response, repeat requests with highest response
    time, cause server to utilize more memory, CPU, disk, etc.\\

    \makecell{Denial of Inventory \\ (\ref{sbf})} & Holds or blocks items from a limited
    inventory but are never purchased or confirmed \\
    \makecell{ Scalping (\ref{sbf})} & Purchases items with limited availability of
    sought-after goods or services (typically high volume and automated) \\
    \makecell{Spamming (\ref{sbf})} &  Abuse of commenting or review system in applications
    to post content with advertisements, photos, etc. \\
    \makecell{Account registrations \\ (\ref{sbf})} & Creates new accounts in bulk. The
    accounts may subsequently be misused  \\
    \makecell{Card cracking (\ref{sbf})} & Brute force attack against payment
    card processes to identify the missing values (start/expiry date, CVV, etc.) \\

    \makecell{Carding (\ref{sbf})} & Validating stolen credit card details against a
    merchant's payment processes to identify valid card details\\

    \makecell{Credential stuffing \\ (\ref{ba})} & Attacker uses a list of valid usernames and
    passwords with the intention to brute force access  \\
    \makecell{Plaintext tokens (\ref{ba})} & Retrieve sensitive authentication details, such as auth tokens
    and passwords in the URL \\



    \hline
    \end{tabularx}
\end{table*}

%% file: sources/design.tex

\begin{figure*}[!ht]
    \centering
        \begin{subfigure}[b]{0.42\textwidth}
            \centering
       \includegraphics[width=\textwidth]{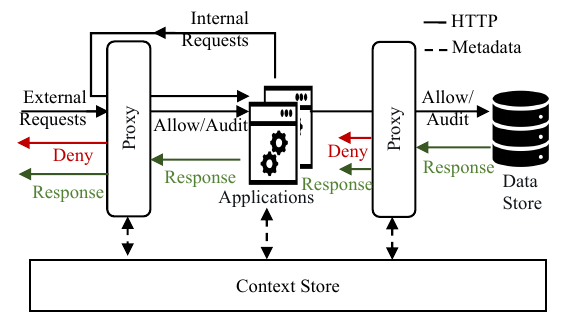}
       \caption{Overview}\label{fig:sysdesign}
         \end{subfigure}
         \hspace{4mm}
        \begin{subfigure}[b]{0.27\textwidth}
            \centering
       \includegraphics[width=\textwidth]{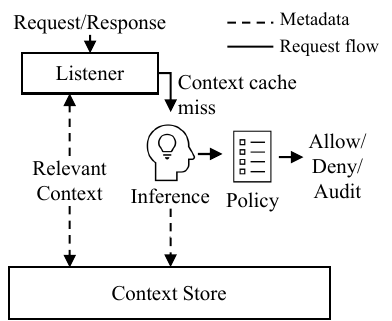}
       \caption{Proxy components}\label{fig:components}
        \end{subfigure}
       \centering 
       \caption{System architecture}\label{fig:arch}
   \end{figure*}
     
\section{System Design \label{sec:design}}

This section describes the design of \sysname{,} the security framework we built
to help cloud administrators.         
A typical web application will accept incoming  HTTP(S) requests and provide an
appropriate response. To construct the appropriate response, the server may
internally make  calls to other services and data stores.  
Figure~\ref{fig:sysdesign} shows the path of an HTTP request in \sysname{.}  
External or internal requests to applications or data stores are intercepted by
a proxy service. The proxy evaluates the configured policies to \emph{allow} the
request, \emph{deny} the request or \emph{audit} the request. Along with
enforcing policies,  proxies also perform out of band logging, independent of
the request paths.

\nip{TLS termination.} Proxies in \sysname{} are deployed based on the API
server design. If communication happens over HTTPS, TLS termination needs to
happen at the gateway or within the application \emph{sidecar}. A sidecar is a
separate container that runs alongside an application container. Sidecar is a
common architecture pattern employed in cloud deployments that act as helpers
for the primarily workload. \sysname{} proxy should be in the path of the 
request after TLS termination for obtaining maximum benefit 
 because \sysname{} rely not only
on url paths, but also on request form data.

       

   

\subsection{Request Tagging\label{sec:tagging}} When HTTP requests arrive at the
proxy (\figref{fig:sysdesign}), \sysname{} utilizes large language models
(LLMs)~\cite{Thrun1998,FloridiC20,Devlin,Brown} to automatically annotate the
request with a predetermined set of tags (classes). \figref{fig:components}
shows the components within the proxy. Once the proxy listener thread picks up a
request, it is parsed and added to a context data store. Specifics of what
context is stored is described later in the section. The request is then
prepared to be sent to a LLM if needed. The resulting request tags and details
are forwarded to the policy enforcement component which allows or denies the
request. The set of tags we use were derived empirically, based on their
effectiveness to enforce a wide array of policies that prevent difficult to
detect attacks on APIs, and have  high success rates when derived using openly
available LLMs. Our design is not dependent on a fixed set of tags, and allow
new tags to be added based on the needs of the policy administrator. Details of
the LLM prompts and models used to derive these tags are explained in
\secref{sec:prompts}.

\sysname{} may assign multiple tags to each request. Tags are organized into two
categories. Business flow tags (\ref{newuser}\ldots\ref{addcart}) and technical
flow tags (\ref{auth}\ldots\ref{login}). Tags in both categories function the same
way and the distinction is only to simplify their description and organization
when used for policy definitions.

\nip{Business tags.} These tags are assigned to requests that match specific 
\emph{business flows}, independent of the underlying applications represented by the APIs.
In this context, we define business flows as tasks that accomplish an 
action meaningful to the application.
There are many business flows common across multiple applications. 
E.g., users logging into the 
API/application or a user adding an item to a shopping cart are 
common patterns. 
Any HTTP request identified as 
a \emph{logging in} flow is then assigned the \emph{login} tag. Policies then check 
 requests with a \emph{login} tag for credential stuffing, 
 sending sensitive authentication details (like passwords) in the URL, etc. 
 The list of tags, their usage details, and the threats they prevent are described next:

\begin{enumerate}[label=\textbf{ B\arabic*},ref=B\arabic*, widest=00, noitemsep]
    \item \label{newuser} \nip{User registration} [registering new users]: This
    tag is useful to check the requests for bulk account creation, credential
    stuffing, credential cracking, empty user registration, sensitive data
    exposure, impersonator bot, massive account registration, etc.
    \item \label{comment} \nip{Commenting} [posting comments, reviews, etc.]:
    This tag is useful to check the requests for social media bot attacks,
    language misuse, spamming, etc.
    \item \label{purchase} \nip{Purchase product} [buying items in e-commerce
    apps]: This tag is useful to check the requests for business logic DoS,
     scalping,  
    preventing legitimate users from buying products, etc.
    \item \label{addcart} \nip{Add to cart} [adding items to shopping carts]:
    This tag is useful to check the requests for denial of inventory, scalping,
    purchase bots, etc.
\end{enumerate} 

\nip{Technical tags.} These tags are assigned to requests that match specific
\emph{technical flows}. We define technical flows as tasks necessary for correct
execution, but by themselves do not  perform actions relevant to core
application business logic. E.g., validating input parameters in a request, limiting
number of records in response based on user request, etc. Similar to business
flows, many technical flows are also common across applications. The list of
technical tags and their usage details are given below:
\begin{enumerate}[label=\textbf{L\arabic*},ref=L\arabic*, widest=00, noitemsep]
    \item \label{rp} \nip{Response data limit}: For requests with parameters
    that control response data size like \lstinline{number of records}. This tag
    is useful to check the requests for exploitation leading to
    denial-of-service (DoS), resource starvation, operational costs increase due
    to higher CPU, network bandwidth etc.
    \item \label{login} \nip{Login} [for user logins]: Login tags are used to 
    check  requests for credential stuffing, attempts to brute force
    passwords, flagging APIs sending sensitive authentication details (like
    passwords) in the URL, etc. 
    \item \label{logout} \nip{Logout} [for user logouts]: Used to
    check requests for session hijacking, testing for session timeout, 
     sensitive data exposure, etc.
    \item \label{upload} \nip{File upload} [uploading files]: This tag is useful
    to check the requests for form spam, large files denial-of-service (DoS),
    skewing etc.

    \item \label{auth} \nip{Contains authorization tokens}: This is useful
    to detect weak token secrets, adding fingerprint information to the token, etc. 
\end{enumerate} 

\subsection{Tag Parameters\label{sec:des:tagparam}}
If we reconsider the prior example of a request influencing the number of 
records in its response ({\RaggedRight \lstinline[language=REST, columns=fullflexible,
breaklines=true]!GET /query?numResults=10!}), an effective cluster 
level policy would be to deny the request if the number or results requested it too high. 
To enforce such a policy, the policy enforcement point needs to know 
how many records the current request is requesting. In this example, 
this is the value of the request parameter \lstinline{numResults}. For 
a different API endpoint this could be value of a different 
request parameter like \lstinline{count}. 
Also such a parameter can be part of the request body, not just the URL. 
All such variations, with the same semantic meaning, makes it difficult 
for policy authors. To prevent this, we extract the 
different parameter values from different request parameters with  
similar meanings and save it as a predefined variable name (policy variable). This 
allows the policy writer to always use the predefined policy variable.
During runtime, we fill policy variables with semantically matching request parameter values.
Few of the request tags, policy variables corresponding to those tags and 
common request parameters that get extracted into those tags are shown in 
\tabref{tab:tagparameters}. From the table, if an incoming API call 
gets tagged as \emph{\Responsedatalimit{,}} then value of any request parameters like
count,  numResults, etc., gets extracted and saved into \lstinline!<num_records>! 
variable. \secref{sec:prompts} describe how we use LLMs to derive these.
\begin{table}[ht]
    \centering
    \begin{threeparttable}
        \caption{\label{tab:tagparameters} Tag parameter names associated with
        different request tags.   }
        \centering
    \begin{tabularx}{\linewidth}{lXX}

    \hline
    Tag &  Policy variables & Common request parameters   \\ \hline
    \Login{}       &   {\lstinline!<username>!}                  & user, username, user\_name, $\ldots$\\ 
    \Fileupload{} &   {\lstinline!<file_name>!}                  & file, file\_name, fileName, $\ldots$\\ 
    \Userregistration{}   & {\lstinline!<username, email>!}      & user, email\_address, email, $\ldots$\\ 
    \Addtocart{} & {\lstinline!<quantity, product_id>!}           & quantity, amount, product, $\ldots$\\ 
    \Purchaseproduct{}   & {\lstinline!<quantity, product_id>!}   & quantity, amount, product ,$\ldots$\\ 
    \Responsedatalimit{} & {\lstinline!<num_records>!}            & count, numResult, num\_records, $\ldots$\\
    \Commenting{}  &   {\lstinline!<comment>!}                    & text, comment,$\ldots$\\ 
    \hline
\end{tabularx}
\end{threeparttable}    
\end{table}

\subsection{Context \label{sec:des:context}} Request tags by themselves do not
provide a enough metadata to enforce all required policies. \sysname{} also
maintains two metadata structures for capturing the \emph{state} of
applications. First one for \emph{per request} attributes (request context) and
a second one for \emph{per container} resource utilization (container context).

\nip{Request context.} For a request $R$, \sysname{}  generates and stores a
5-tuple context: $$R_c = \langle ts, src, dest, hist  \rangle \text{,
where,}$$ 
\mathnip{$\mathbf{ts} \gets \textbf{time stamp of request.}$}  HTTP by itself,
    does not carry a time stamp attribute. Time stamp $t$ of a request $R$ is
    calculated as the time when the proxy first received the HTTP request. 

\mathnip{$\mathbf{src} \gets \textbf{source attributes.} $} 
    \sysname{} finds the client IP address by examining the
    underlying TCP connection. Common HTTP headers like \lstinline{X-Forwarded-For} or
    \lstinline{X-Real-Ip} are added to the map of source attributes if present.

\mathnip{$\mathbf{dest} \gets \textbf{destination attributes.} $} To retrieve the
    destination attributes ($d$), the HTTP request is separated into its distinct
    components (host, port, path, query string, headers and body) and stored. 
    Further, based on the destination host and port, we retrieve the identity of
    the container 
    hosting the API server which will serve request $R$. 

\mathnip{$\mathbf{hist} \gets \textbf{past tags assigned to request.} $} Business or 
technical tags  assigned to past requests with the same API path by \sysname{} are saved ($hist$)
for performance and summary analysis performed during policy evaluation.

\nip{Container context.} Container context is extracted from the control plane
of the cloud orchestration platforms managing the cloud cluster.
 Popular platforms like Kubernetes~\cite{k8},
OpenShift~\cite{openshift}, etc., come pre-installed with, or can be easily
configured with tools to emit system metrics related to memory, CPU and IO
usage. \sysname{} queries the cloud control plane to get the active deployments
and running containers. For each of these containers we query system metrics to
record the context tuple. For every running container $C$ in the cluster, \sysname{}
maintains a 4-tuple container context row $C_c$, given by 
     $$C_c = \langle ts, m, c, i  \rangle \text{, where,} $$ 

\mathnip{$\mathbf{ts} \gets \textbf{time stamp.}$}  Attribute $t$ contains the time stamp 
at which the specific container context was generated.

\mathnip{$\mathbf{m,c,i \gets \textbf{memory, CPU, IO.} }$} Attributes $m$, $c$ and $i$  refers to memory usage, 
CPU usage and IO usage respectively of container $C$ at time $t$.
Container context snapshots are created every fixed time intervals. Currently 
in \sysname{} we have set this time interval to be one minute.
Policy definitions combine request tags with request and container context to author 
protections against common API vulnerabilities.
Context metadata along with request tags, 
provide a robust framework for defining and enforcing policies that can effectively
tackle~\ref{rc},~\ref{sbf} and~\ref{ba}  defined in \secref{sec:introduction}.

\subsection{Caching\label{sec:des:caching}}
For each HTTP request, we extract a unique key from the request method (POST,
GET, etc.), and url path (excluding parameters). The reply from the LLM is
parsed and cached with the generated unique key. The cache entry contains 
the request tag and corresponding tag parameters. Tag parameter \emph{values} 
are not cached as they change every request.
 For every new request, the
cache is checked for an existing key, and if present, the request tag and tag
parameters are identified from the cache. Once the tag parameters are
identified, tag parameter values are extracted directly from the current
request's body. 

%% file: sources/detail.tex
\section{Policy Definitions\label{sec:policydef}} 

In this section we discuss how request tags and context help make 
 policy definitions effective and easy. 
 Request tags, request context and container context 
 generated by \sysname{} are utilized by policies that decide 
 if the request should be allowed or denied. 
 \sysname{} provides a policy interface that can be implemented 
 by policy authors. Few of the methods in the interface are shown in 
 \lstref{lst:interface}. All tags and tag parameters 
 (\lstinline!tag_detail! in~\ref{lst:interface}), 
 along with request and container context (\lstinline{context} 
 in~\ref{lst:interface}) are passed into the 
policy definition functions by the framework. Policy writers implement each  
 tag specific methods. Framework logic calls the \lstinline!PreTagPolicy! function (line~\ref{lst:line:prepol}) 
 before any tag specific function calls and the \lstinline!PostTagPolicy! function (line~\ref{lst:line:postpol}) after all 
tag specific policies. These are useful to define tag independent policies 
as well as cross tag policies.

\begin{lstlisting}[float, language=go2, caption={\sysname{} policy definition interface}, label={lst:interface}, escapechar=|,belowskip=-0.9 \baselineskip]
type Policy interface {
PreTagPolicy(*tag_detail, *context) (bool, error) |\label{lst:line:prepol}|
// Response data limit policy function 
ResponseSize(*tag_detail, *context) (bool, error)
//  Add to cart policy function
AddToCart(*tag_detail, *context) (bool, error)
...
PostTagPolicy(*tags, *context) (bool, error) |\label{lst:line:postpol}|
}
\end{lstlisting}

We organize the broad spectrum of policies that can be written and enforced
using \sysname{} into  three categories, aligned with the security risks
outlined in \secref{sec:introduction}. Though \sysname{} is not limited to
securing only against these categories of risks, we believe they showcase a good
breadth of possibilities available to a cloud security administrator.

\subsection{Unrestricted resource consumption (\ref{rc})} 
An internal or external API call, results in backend servers utilizing resources
like CPU, memory, network, IO, etc. Malicious actors exploit this by invoking
APIs that retrieve larger than expected amounts of data, storing large data,
wasting CPU cycles, etc. This can result in denial of service (DoS) for
legitimate users. If the servers are deployed in the cloud, such usage can incur
additional expenses as well. Ideally an API developer would be diligent and
check against such attacks by limiting resource usage for each request. But,
developers  often do not  set appropriate limits because, limits can be
dependant on the runtime environment, current threat landscape, or simple
oversight.  This problem is exacerbated by fast releases and lack of developer
time in many software development processes. Using \sysname{}, system
administrators can craft policies to prevent attacks causing such unrestricted
resource consumption.

\nip{Policy.} Consider the two APIs below that respond with a certain number of
    records, based on a parameter in the HTTP request:
    \begin{enumerate}[label=\textbf{Req.\arabic*},ref=Req.\arabic*,align=left]
        \item \label{reqone}  {\RaggedRight \lstinline[language=REST,breaklines=true]!GET /feed/list?count=10!. }
        \item \label{reqtwo} {\RaggedRight \lstinline[language=REST,breaklines=true]!GET /commentThreads?part=7&maxResults=30!. }
    
    \end{enumerate}
     These two API requests are part of two independent real world applications.
    Now consider a cloud security
    administrator wanting to set a cluster wide, maximum limit for the number of records 
    any application can return.  Such a need could arise as a temporary measure to limit the impact of an
    ongoing attack or a secondary line of defense in case the API developer
    missed to set limits themselves. To solve this, the administrator can use \sysname{,} 
    and write the policy shown in \lstref{lst:policyimpl}. In this example, \sysname{}
    detects that the request contains an HTTP query parameter and adds the tag \lstinline{Response data limit} (\ref{rp}). 
    Then \sysname{} extracts the parameter value ($10$ for \ref{reqone} and $30$ for \ref{reqtwo}) 
    and saves it into the predefined parameter key for \lstinline{Response data limit}, 
    \lstinline{num_records} (\tabref{tab:tagparameters}). 
    \lstref{lst:policyimpl}, line~\ref{lst:line:num_recs_check} simply checks if this extracted 
    parameter is within a threshold configured by the cloud administrator.

\begin{lstlisting}[float, language=go2, caption={Pseudocode of policy to limit number of records in a response irrespective of the application or request parameter name}, label={lst:policyimpl}, escapechar=|,belowskip=-0.9\baselineskip]
...
ResponseSize(*tag_detail, *context) (bool, error) {
    num_records = tag_detail.num_records 
    // Check if number of records returned is within threshold
    if (num_records < config.record_threshold)  |\label{lst:line:num_recs_check}|
        return true
    else
        return false
}
...
\end{lstlisting}

Similar to request parameter based policies, thresholds on server memory, 
network and CPU usage can be enforced by checking the context metadata 
passed into the policy. Note that container context (\secref{sec:des:context})
reflect the current state of resource usage and is not a prediction of 
how much resources the current API request will consume.
Changes in resource usage from past time points can be probabilistically 
 correlated with past APIs in the same time window to identify APIs
 that repeatedly cause spike resource usage. \sysname{} benefits  
 from similar offline analysis~\cite{DBLP:conf/raid/ValdesS01} by including
 results of these analysis as part of context data.

\subsection{Unrestricted access to sensitive business flows (\ref{sbf})} API
endpoints expose different functionalities, each representing different business
flows in an underlying application. Examples of such flows are: purchasing a
product, making a reservation, adding a comment or review, registering new
users, etc. For some business flows, it is important that access is restricted
based on content or rate of request. E.g., for \emph{purchasing a product} flow,
the number of purchases in a given time window as well as the quantity 
of items in a single purchase is significant. An attacker can
repeatedly purchase all stock of a popular item to resell (scalping). 
Having the ability to define application agnostic policies that prevent such 
actions, across all APIs served from a cluster, is a very valuable tool for a 
cloud system administrator. This is easily implemented in  \sysname{,} 
with a policy shown in \lstref{lst:policypur}. 

As described in \secref{sec:design}, \sysname{} first tags the request
 as \lstinline{Purchase product} (~\ref{purchase}). Then the \lstinline{purchase quantity},
 which is often part of the request json body, is automatically extracted by the
 framework and saved into the predefined parameter key  \lstinline{quantity}
 (line~\ref{lst:line:qtycur}).  
Further, past history of the same request ($R_c$) is accessed and intent
parameter \lstinline{quantity}, is aggregated  (line~\ref{lst:line:qtyaggr})
over the last five minutes (line~\ref{lst:line:qtyaggr}) for products with the
same \lstinline{product_id} as the current request
(line~\ref{lst:line:qtyhaving}).  
This aggregated sum gives total purchases of the product over a five minute
window and the administrator then checks if the sum is within a configured
threshold (line~\ref{lst:line:maxpurchase}).

\begin{lstlisting}[float, language=go2, caption={Pseudocode of policy to limit number of items purchased over a 5 minute time window}, label={lst:policypur}, escapechar=| ,belowskip=-0.9 \baselineskip]
...
PurchaseProduct(*tag_detail, *context) (bool, error) {
    //Number of items purchased in this request         
    quantity = tag_detail.quantity |\label{lst:line:qtycur}|
    cur_product = tag_detail.product_id
    //Total qty in the past 5 minutes
    total_qty = quantity+stream(context.Rc.hist)|\label{lst:line:qtyhist}|
        .window(Time.minutes(5)).aggregate(quantity)|\label{lst:line:qtyaggr}|
        .having(product_id = cur_product)|\label{lst:line:qtyhaving}|
    if (total_qty < config.max_purchase_qty)|\label{lst:line:maxpurchase}|
        return true
    else
        return false
}
...
\end{lstlisting}

\subsection{Broken authentication (\ref{ba}). } Attackers may target an API's
authentication mechanism in order to gain control of other user's account in the
system. To do this, attackers may repeatedly try different passwords, apply
leaked tokens, etc. Since \sysname{}  assigns request tags (e.g., login) automatically 
policy writers can simply check the number of occurrence of requests 
with a specific tag in a given time window as shown in \lstref{lst:policypur}. 
Additional grouping based on source of the request can also be done.

%% file: sources/prompt_tuning.tex


\begin{figure*}[ht]
    \centering
            \includegraphics[width=.7\linewidth]{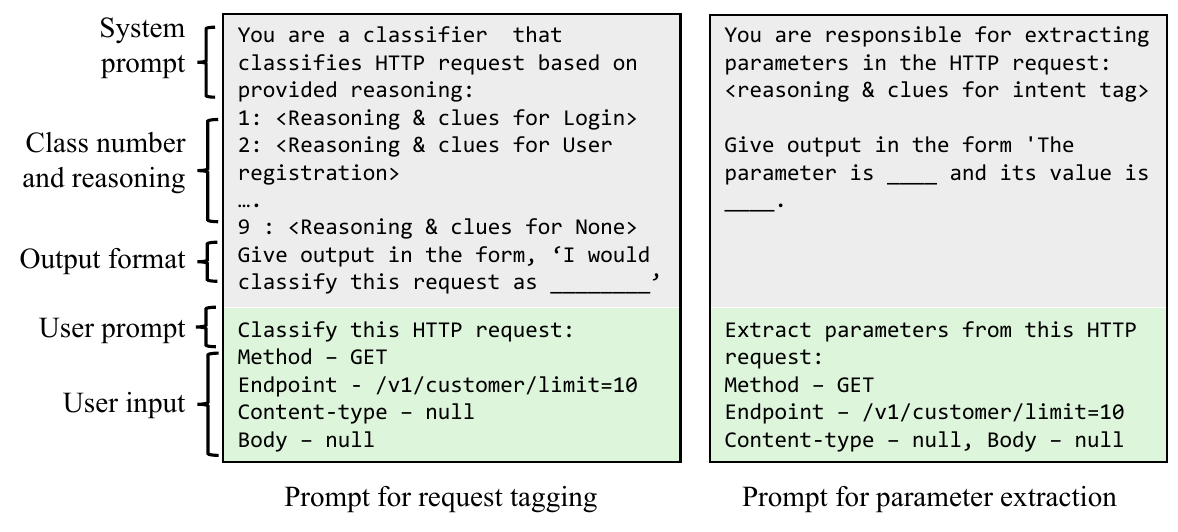}
       \caption{LLM prompts used by \sysname{}. Labels show different sub-components of the prompt.}\label{fig:prompt_format}
   \end{figure*}

\section{LLM Prompts for Request Tagging\label{sec:prompts}}

In this section we describe how \sysname{} derive request tags from REST APIs.
As outlined in \secref{sec:tagging}, every request gets associated with, a
predefined set of tags summarizing the intent of the request, and relevant
parameters for the tags. To derive these tags and tag parameters, we use
specialized prompts to classify the request using a large language model (LLM).

\subsection{Prompt Format}

Generally, specifics of the prompt depend on the LLM being used. Though
configurable, currently \sysname{} uses
\lstinline{llama-2-70b-chat}~\cite{llamamodel} and the corresponding prompt
structures are shown in \figref{fig:prompt_format}. This prompt consists of the
following six components:

\nip{System prompt.} The system prompt sets the overall context, specifying what is
required of the LLM in the given task.

\nip{Class identification number.}   A number assigned to each request tag or
classification. Using a \emph{class number} instead of an actual \emph{class name} here, helps
prevent biases about the name in the decision. We want the LLM to only use the
reasoning and clues we provide for each class, and not rely on the name of the class
itself. In \figref{fig:prompt_format}, the class identification number are 
{\ttfamily\small 1, 2, ..., 9}. 

\nip{Class identification reasoning.} For each class we provide a primary
reasoning and additional clues. The primary reasoning provides the basic
criteria for identifying the appropriate class (tag) for a request. The
clues are provided to give additional information or context for that
classification. The clues could be keywords, phrases etc., that can help
identify a certain class. Adding clues help reduce false positives
substantially. E.g., for the \lstinline{File Upload} tag, the reasoning and
clue we used is:  
{\RaggedRight\color{mycodeblue}\ttfamily\small If the HTTP request uploads
or adds an image/file to the application/server.} {\RaggedRight
\color{mygreen}\ttfamily\small The request should have file name or
reference to} {\color{mygreen}\ttfamily\small file being uploaded.} 
Here, the first part of the above text ({\ttfamily\small If...server.}, in
blue) is the primary reasoning and the next sentence ({\ttfamily\small
The...uploaded.}, in green) is the clue.

\nip{Expected Output Format.}  Next, we specify the format in which output is expected.
    This forces the LLM to provide a result in a pattern we expect, making it easier to write 
    a parser to extract the predicted classes from the output. 

\nip{User prompt.} Now we specify a prompt prefix to let the 
    LLM know that all the instructions specified prior to this line is to be applied 
    to the following input.

\nip{User input.} Finally, the incoming HTTP request to be classified is specified. 
The request is provided in a text format. A typical example of HTTP request is shown 
in \figref{fig:prompt_format}.

Finally, we provide an additional 
\lstinline{None} category, representing requests that does not need any tags. 
In general we observed better results when prompting the LLM to \emph{do this,}
instead of prompting \emph{do not do this}. By prompting the LLM to assign \lstinline{None} category instead of prompting to not assign any tags (if request does not match any flows) was more effective.

\begin{figure}[ht]
    \centering
        \begin{subfigure}[b]{0.6\linewidth}
            \centering
            \includegraphics[width=\linewidth]{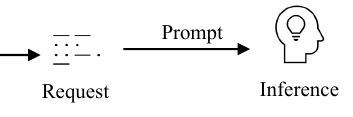}
            \caption{Single, multi-class prompt mode}\label{fig:multi_mode}
        \end{subfigure}
        \begin{subfigure}[b]{0.6\linewidth}
            \centering
            \includegraphics[width=\linewidth]{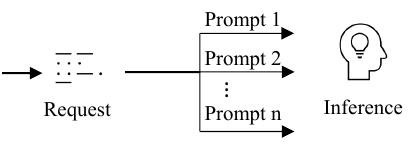}
            \caption{Parallel single-class prompt mode}\label{fig:binary_mode}
         \end{subfigure}
       \caption{\sysname{} prompting modes}\label{fig:prompting_modes}
   \end{figure}

\subsection{Prompting Modes}

\sysname{} supports two configurable modes to send the framework's prompt to the
LLM for inference as shown in \figref{fig:arch}. \emph{Single mode}, uses one
curated prompt to assign all the tags to a given request. This is a multi-class
classification prompt. \emph{Parallel mode}, uses one prompt for each tag and
creates multiple inference requests in parallel. The prompt format remains the
same for both modes, except, in parallel mode, each  prompt  will only contain
two fixed class identification numbers and corresponding class identification
reasoning. One for the tag class and one the \lstinline{None} class. The primary
advantage of parallel mode is the ability to add new tags easily. We observed
that while operating in single mode, using multi-class prompts, adding a new tag
and corresponding class identification number with reasoning would often disrupt
the accuracy of already configured tags.

%% file: sources/implementation.tex
\section{Implementation\label{sec:impl}} 
 We implemented a functioning prototype of \sysname{} using Go~\cite{golang}, 
 Envoy~\cite{envoyproxydoc} and WebAssembly~\cite{wasmdoc}. In this section,
 we describe this prototype.

\nip{Envoy proxy.} Envoy is a layer-7 proxy and communication bus~\cite{envoyproxydoc}. Envoy allow users to add \emph{filters}, which customize 
what Envoy does after intercepting an HTTP request. 
We use Envoy as the proxy shown in \figref{fig:sysdesign}. 
We build a custom plugin filter for envoy and configured Envoy to route requests through our plugin. 
The functional components shown in \figref{fig:components} are 
implemented within this plugin. 

\nip{WebAssembly.} Our custom plugin use the WebAssembly~\cite{wasmdoc} (Wasm) binary format. 
The WebAssembly language specification describe a memory-safe, stack-based, sand-boxed execution environment aiming to execute code at native speed. 
Envoy embeds a WebAssembly engine within it, which loads and runs the custom plugin we built. We implemented our request tagging logic in the plugin with Go, which we then compile into a Wasm module.  

\nip{Custom plugin.} The plugin implementation contains the prompts we described in \secref{sec:prompts}. The incoming request is added to the 
prompt (user input part of the prompt) and forwarded to the configured LLM 
module for inference. The inference callback function parses the reply 
from the LLM and assigns the tag and fills the request context with the 
appropriate tag parameters. 


\renewcommand\tabularxcolumn[1]{m{#1}}

\begin{table*}[ht]
    \centering
    \begin{threeparttable}

    \caption{\label{tab:APIdataset} Dataset details}
    \centering

    \begin{tabularx}{\linewidth}{llX}
    \hline
    Dataset & \#APIs~ & Associated Applications (\#APIs) \\ \hline
    \Dsone{} & 501 & API-Football (85), SportScore (73), Imgur (73),
    Tiktok$^\alpha$ (103),  Instagram$^\beta$ (63), Twitter (16), Webcams (13), Youtube$^\delta$ (18)
    Deezer (10),   Pinnacle Odds (8), BraveNewCoin (7), OpenAPI (6),  BetsAPI (6), Telize (5), 
    MapTiles (4),  COVID-19 (3), IP Geo Location (2),  Aka (2), Rapid
    Translate Multi Traduction (2),  Google Search (2) \\

    \Dstwo{} & 915 & PayPal (124), Square (177), Apache Fineract (614) \\
    \Dsthree{} & 1026 & Shopify (313), Shopizer (279), OWASP Juiceshop (434) \\
    \hline
    \end{tabularx}
    \begin{tablenotes}[para, flushleft]
        \small
        \item $^\alpha$ Consists of TokApi–mobile version, Tiktok video and TikTok All in One. $^\beta$ Consists of Instagram Bulk Profile Scrapper, Instagram Scraper and Instagram.  \item $^\delta$ Consists of Youtube v3, Youtube Search, and Youtube Download.
      \end{tablenotes}  
    \end{threeparttable}          
\end{table*}

\nip{Policy enforcement.} After the appropriate tags are identified, implementations of the interface 
functions as described in \secref{sec:policydef} are invoked. If any of the 
policy functions return false, the plugin replies with an HTTP error 403.
Otherwise the request is forwarded to its original destination.

%% file: sources/evaluation.tex
\section{Evaluation\label{sec:eval}} 

In this section we evaluate the effectiveness of 
 \sysname{} as a framework for API security. We look at 
 the ability of our system to
correctly identify and reliably enforce policies.
Specifically, we measure the
effectiveness of \sysname{} based on the following criteria:

\begin{itemize}\label{criteria}
    \item \textbf{Tag popularity}: One of our core assumption is that 
    business and technical flows (consequently request tags) will repeat across applications. 
    We evaluate this hypothesis by analyzing the frequency with which specific tags 
    appear in common application APIs. More popular the tag, more generalizable the 
    framework will be across applications.
    \item \textbf{Tag association accuracy}: In order to identify the correct
    policies to apply on a request, \sysname{} needs to associate the correct
    tag(s) with every request. This is same as the classification accuracy of
    our inference step (\figref{fig:components}). Higher the classification
    accuracy, higher the tag association, and our confidence that the policies
    defined for a specific flow is being invoked and enforced.
    \item \textbf{Performance}: Finally, we measure the latency incurred by \sysname{}
    to perform policy enforcement.
\end{itemize}

\subsection{Dataset Design\label{sec:datasets}}
Commonly used HTTP datasets like PKDD2007~\cite{pkdd2007} and
CSIC2010~\cite{gimenez2010http} are more than 14 years old and contains
non-English characters and words in request URL and body. Since our tagging
relies on semantics and not just syntax of the requests, we find them
inappropriate for our evaluation. Instead, we curated three new datasets with
different characteristics as described below:

\nip{\Dsone{}}: We selected top 25 most popular application APIs for
    2023, as identified by RapidAPI~\cite{RapidAPI}. This list includes 501 APIs
    from TikTok, Instagram, Google, and others. Full list of applications 
     are shown in
    \tabref{tab:APIdataset}. This dataset consists of APIs from diverse domains
    like social media, news and statistics related to sports, video streaming 
    etc. We consider this diversity good for testing the generalizibility of \sysname{} across applications.

 \nip{\Dstwo{}}: This dataset was created by selecting APIs specific to the
    financial domain. This includes  PayPal~\cite{paypal}, Square~\cite{square},
    and Apache Fineract~\cite{apachefineract}. This dataset contains 915 API
    calls. Since financial applications will have many of the business and
    technical flows described in~\ref{sec:tagging}, this will be a good dataset
    to measure tag classification accuracy of \sysname{}.

 \nip{\Dsthree{}}: This dataset consists of API calls from popular e-commerce
    applications like Shopify~\cite{shopify}, Shopizer~\cite{shopizer} and test
    applications like OWASP Juice Shop~\cite{owaspjs}.  There are 1026 API calls in 
    this dataset. E-commerce applications have most
    of the identified business and technical flows and will serve to ensure our
    results are generic beyond a single domain. 

    We labelled all APIs in the dataset manually with their request tags (ground
    truth) based on API documentation. More details of these APIs and their per
    application breakdown are given in \tabref{tab:APIdataset}.

    \begin{figure*}[!ht]
        \includegraphics[width=\linewidth]{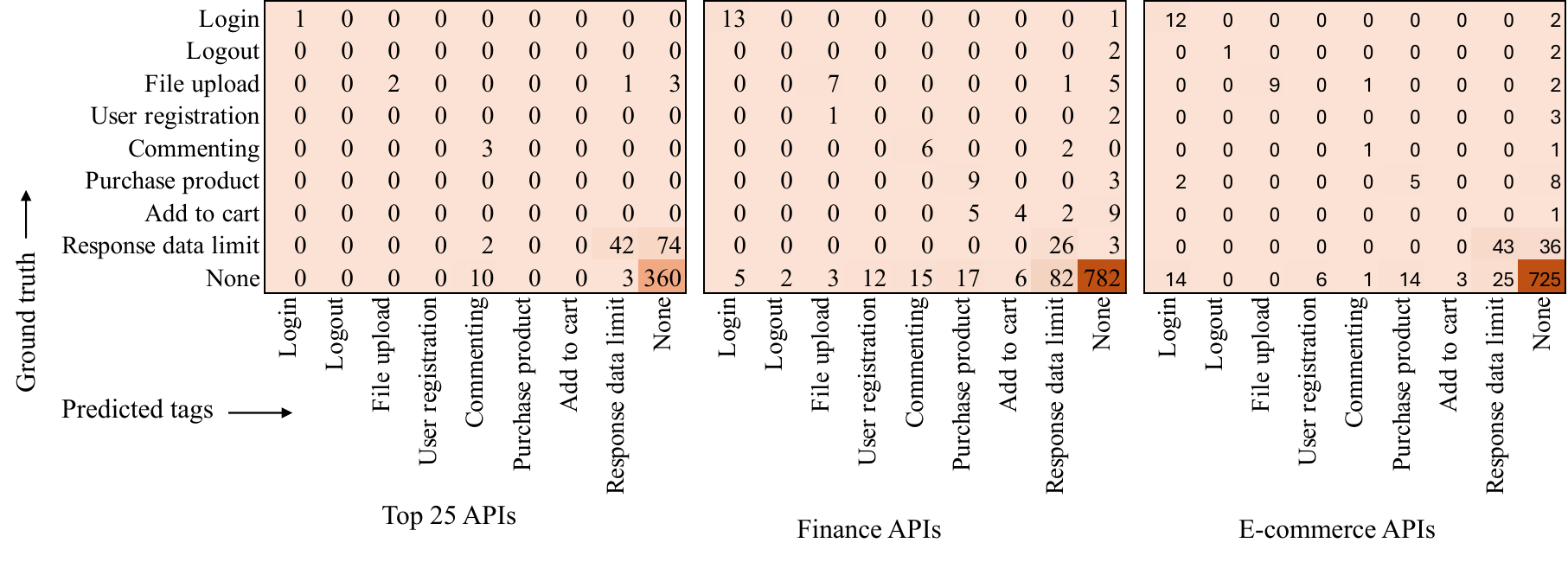}
           \caption{Single mode, confusion matrix for datasets}
        \label{fig:multiclass}
       \end{figure*}
    
    \begin{table}[ht]
        \centering
        \begin{threeparttable}
        \caption{\label{tab:tagscount}Tag popularity measured across datasets}
        \centering
        \begin{tabular}{llll}
        \hline
        Request tags & $D1^\alpha$ & $D2^\beta$ & $D3^\gamma$ \\ \hline
        \Login{} & 1 & 14 & 13 \\ 
        \Logout{} & 0 & 1 & 2 \\ 
        \Fileupload{} & 6 & 12 & 13 \\ 
        \Userregistration{} & 0 & 3 & 3 \\ 
        \Commenting{} & 3 & 2 & 8 \\ 
        \Purchaseproduct{} & 0 & 15 & 13 \\ 
        \Addtocart{} & 0 & 1 & 20 \\ 
        \Responsedatalimit{} & 118 & 79 & 29 \\
        \Containsauthorizationtokens{} & 75 & 275 & 181 \\ \hline
        \end{tabular}
        \begin{tablenotes}
            \small
            \item $^\alpha$ \#Requests in \Dsone{} with Tag
            \item $^\beta$  \#Requests in \Dstwo{} with Tag
            \item $^\gamma$ \#Requests in \Dsthree{} with Tag
    
          \end{tablenotes}
        \end{threeparttable}    
        \end{table}

\subsection{Tag Popularity}
We first analyze the popularity (frequency of occurrence) of request tags in
common application APIs. We analyzed the three data sets listed in
\secref{sec:datasets}, and the results are summarized in \tabref{tab:tagscount}.
In summary, $26\%$ of all requests in the top 25 APIs in 2023 belong to 
one of the flows we identified (other than \lstinline{None}).
Similarly $10\%$ of requests in \Dstwo{} and $14\%$ of requests in \Dsthree{}
 belong to at least one of the identified flows. Unsurprisingly,
  \Dsone{} primary consists of APIs that query and filter data (e.g., from 
 TikTok, Instagram, etc.), with only few business flows. Even in this 
 scenario we see good popularity for technical flows like \containsauthorizationtokens{,} and \responsedatalimit{.}

\subsection{Tag Association Accuracy}
\sysname{} needs to associate the correct tag(s) with 
    every request in order to identify the correct policies to be enforced. 
    We now show the  accuracy of this association. 
We used llama-2-70b-chat~\cite{touvron2023llama} model with default parameters and maximum token size set to 20, to perform zero-shot classification 
of requests as belonging to the predefined
 business and technical flows (\secref{sec:design}). 
Then, we compared the ground truths  against the results from the LLM, 
for both operation modes.

\nip{\Modeone{.}} Tag association accuracy for single, multi-class prompt mode
is summarized in \figref{fig:multiclass}.
We have overall classification accuracy of $81.2\%$, $85.2\%$ and $82.9\%$ for
datasets \Dsone{,} \Dstwo{,} and \Dsthree{} respectively. We observed higher 
rates of false
positives with requests being tagged with \responsedatalimit{,}  
 when they are not supposed to. In practice, we are able 
to work around few of the false positives during tag parameter extraction stage. 
E.g., if a request is tagged as \responsedatalimit{,} we expect a corresponding 
tag parameter. If the tag parameter is missing we can often identify the assigned 
tag as a false positive. 

 \nip{\Modetwo{.}} Tag association accuracy for \modetwo{} is summarized
 in Figure~\ref{fig:parallel_mode_tag_assoc}.
 For \Dsone{,} overall accuracy is very high (96\%) and false positive rate is very low (3\%).
 We saw highest false positives for \Dstwo{} (27\%, overall) with login,  logout and 
 user registration tags being assigned wrongly most.  For \Dsthree{}, we have 
 an accuracy rate of 82\%. We believe as LLM models improve, or by using 
 LLM models specifically built or tuned for security, our accuracy 
 rates will become higher. 
Overall, in our prototype, \sysname{} is able to assign correct request tags for 82\% of the APIs we 
evaluated. 

\begin{table}[htbp]
    \centering
    \caption{Parallel prompt tag association accuracy\label{fig:parallel_mode_tag_assoc}}
      \begin{tabular}{lrrrrrr}
            \hline
            Dataset $\rightarrow $ & \multicolumn{2}{c}{Top 25 APIs}  &    \multicolumn{2}{c}{Finance APIs} &     \multicolumn{2}{c}{E-com. APIs} \\
            \hline 
            Request tags& \multicolumn{1}{r}{FPR$^\alpha$ } & \multicolumn{1}{r}{TPR$^\beta$ } & \multicolumn{1}{r}{FPR} & \multicolumn{1}{r}{TPR} & \multicolumn{1}{r}{FPR} & \multicolumn{1}{r}{TPR} \\
            \hline 
      Login & 5.6   & 100   & 28.8  & 100   & 19.8  & 92.3 \\
      Logout & 4     & \multicolumn{1}{r}{-} & 42.2  & 100   & 25.8  & 0 \\
      File upload & 2     & 100   & 37.9  & 100   & 20.9  & 92.3 \\
      User registration & 1.4   & \multicolumn{1}{r}{-} & 32.5  & 100   & 15.7  & 66.7 \\
      Commenting & 1.4   & 100   & 20.7  & 100   & 14.5  & 75 \\
      Purchase product & 0.2   & \multicolumn{1}{r}{-} & 21.4  & 57.1  & 16.2  & 61.5 \\
      Add to cart & 0.8   & \multicolumn{1}{r}{-} & 20.4  & 0     & 15.8  & 60 \\
      Resp data limit & 11.7  & 97.458 & 12.4  & 98.7  & 17.2  & 96.6 \\
      Total & 3.1   & 97.656 & 27.2  & 93.7  & 18.3  & 79.2 \\
      \hline
      \end{tabular}%
     \begin{tablenotes}
        \small
        \item $^\alpha$ False positive rate (FPR) $ \gets \frac{\text{false positives}}{\text{false positives} + \text{true negatives}} $
        \item $^\beta$  True positive rate (TPR) $ \gets  \frac{\text{true positives}}{\text{true positives} + \text{false negatives}}$

      \end{tablenotes}

  \end{table}%

\begin{figure}[!ht]
    \includegraphics[width=.8\linewidth]{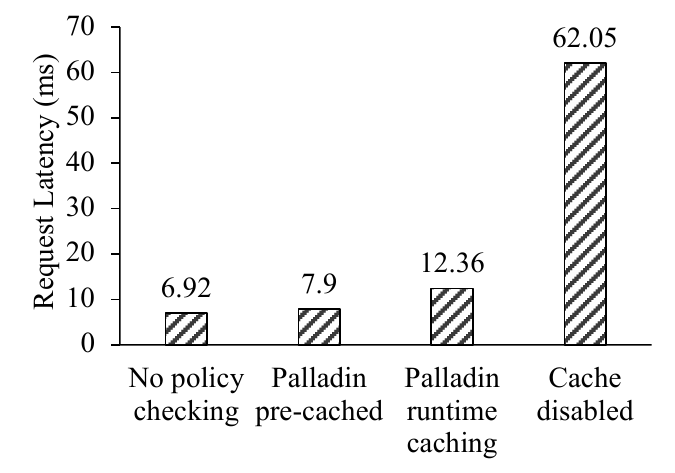}
       \caption{Average response latency of requests in top-25 APIs. No policy checking 
       shows the baseline without \sysname{.}
       \sysname{} default pre-caches LLM responses. Runtime caching starts with an empty cache and populates cache on every new inference. }
    \label{fig:latency}
   \end{figure}

   \subsection{Latency}

   Finally we look at how \sysname{} impacts the response time of API requests. 
   The test server 
   accepting incoming requests was hosted in an Intel Xeon server (12 core, 64 GB RAM).
   \figref{fig:latency} 
   show the average response time of API requests in \dsone{} data set.
   The test server hosts an API stub to respond to the incoming requests.
   LLM inference is performed by an external service. 5000 API requests were 
   randomly picked from \dsone{} for this evaluation.
   The baseline run
   without \sysname{} intercepting the requests or enforcing policies is shown
   as \lstinline{No policy checking} in the Figure. \sysname{} with requests pre-cached 
    operates with $14\%$ additional latency when compared to the baseline. 
    When requests are not pre-cached, 
    first cache miss will trigger tag generation, but this 
    cost is amortized over the total run until API endpoints change. As the \#requests 
    increase and system runs for a long time, average response time in
     runtime caching scenario 
    will get closer to pre-cached response time. An extreme example, when caching is fully 
    disabled, is also shown for completeness where every request incurs an inference call.

%% file: sources/related.tex
\section{Related Work\label{sec:related}} 

Application firewalls and intrusion detection systems (IDS) are the most commonly used  
defense mechanisms against attacks on APIs. 
Firewalls filter and monitor HTTP traffic to and from an application,
in order to protect against SQL injection, cross-site scripting,
cross-site forgery and other attacks. A firewall does this by blocking network
traffic from certain IPs, enabling traffic only on certain ports and, 
utilizing signature-based attack detection techniques and more. 
Attack detection as done by such firewalls 
is a well studied topic. 
Popular attack detection systems like Bro~\cite{PAXSON19992435} and 
Snort~\cite{DBLP:conf/lisa/Roesch99} rely on  attack signatures  while 
others~\cite{networkintrusion1,networkintrusion2,networkintrusion3} 
rely on payload analysis. 
Typically, they rely on a set of rules written
by the administrator, who therefore must have an in-depth
knowledge of the applications to be protected.
Anomaly detection systems~\cite{intrusiondetection2,DBLP:conf/sp/Denning86,DBLP:conf/id/NeumannP99} 
can detect new and unknown attacks but  
may also produce a relatively higher number of false positives.
For unlabeled network anomaly detection, Portnoy et
al.~\cite{intrusiondetection1} presented a clustering algorithm to discover and
separate outliers in the training dataset. The clusters of normal data are
then used to construct a supervised detection model.
Our system is not a replacement for existing 
firewalls or IDS tools, but an additional tool to enhance the 
overall security of the cluster.
 
\nip{API security and layer-7 attack detection.}
Another option to secure web applications is by verifying security properties of
applications~\cite{DBLP:conf/csfw/AkhaweBLMS10} and
protocols~\cite{DBLP:conf/wsfm/BhargavanFG06,DBLP:journals/fac/GordonP05} using
a variety of tools~\cite{DBLP:conf/csfw/Song99}. 
Forcehttps~\cite{DBLP:conf/www/JacksonB08} enforces stricter error processing by
the browser when web application injects a specialized cookie. Such stricter
processing can protect the URL parameters, and secure cookies from attackers and
users who click through security warnings. 

Prior work that incorporates
application layer syntax for anomaly detection~\cite{ApplayerAnomalyDetection}
includes TokDoc~\cite{tokdoc}, which intelligently replaces suspicious parts in
a requests by benign data the system has seen in the past. Oza et
al.~\cite{OZA2014242}, described a byte level n-gram analysis to protect against
HTTP attacks. HMMPayl~\cite{ARIU2011221} uses Hidden Markov Models to analyze
HTTP payloads to protect against  XSS and SQL-Injection attacks. Though these
systems are effective against specific attacks, they are not designed to be a
mechanism to control attacks like unrestricted access to sensitive business
flows or unrestricted resource consumption.

\nip{Request classification.}
Approaches to detect known and unknown web application attacks by analyzing HTTP
requests with and without machine learning  
has been explored before as well. Machine learning techniques were shown to
improve the detection capabilities of the web application
firewalls~\cite{DBLP:conf/icmla/BetartePM18}. Utilizing the structure of HTTP
queries that contain parameters and comparing it to pre-established profiles was
shown to be effective for anomaly detection~\cite{DBLP:conf/ccs/KrugelV03}.
 The primary goal for these studies usually is to classify an
HTTP request as normal or abnormal. Li et al.~\cite{9154702} specify using
Word2vec paragraph vectors to build an efficient anomaly detection model. The
focus these works is to find anomalous HTTP requests and orthogonal to our needs
to provide a runtime policy enforcement system.

%% file: sources/conclusion.tex
\section{Future Work and Conclusion\label{sec:conclusion}} 

We are currently looking into utilizing LLMs built specifically 
for cyber security~\cite{motlagh2024large}. LLMs trained specifically
on security text data, including security and network logs, security literature, etc., can better tag API requests with security specific 
flows. 
We are also looking at  offering automated 
corrections to potential risks by identifying requests, where, instead of denying the request, we can modify the request, making it safer, without affecting 
the semantics expected by the client. 

 In this paper, we introduced a new framework called \sysname{} to augment existing security control mechanisms in the cloud to prevent layer-7 attacks 
 on APIs. This framework utilizes large language models to successfully
 group similar requests and automatically extract relevant 
 information from these requests to make policy definitions simple.
 \sysname{} also combines request and environment context with 
 the intent of the request,  making policies effective. 
 With examples and evaluations, the paper shows how to use the framework to 
 prevent common threats against APIs.